\documentclass[11pt]{article}

\usepackage{booktabs}       
\usepackage{tabularx}
\usepackage{multirow}       
\usepackage{threeparttable} 
\usepackage{CJKutf8}
\usepackage{amsmath}
\usepackage{tikz}
\usepackage{float}
\usetikzlibrary{arrows.meta, positioning, fit}
\usepackage{listings}
\usepackage[most]{tcolorbox}
\usepackage{algorithm}
\usepackage{algpseudocode}
\usepackage{pdflscape}
\usepackage{multicol}
\usepackage{listings}
\usepackage{caption}
\usepackage{booktabs}
\usepackage{tabularx}
\usepackage{ragged2e}
\usepackage{makecell}
\usepackage{array}
\usepackage{adjustbox}

\newcolumntype{Y}{>{\RaggedRight\arraybackslash}X} 

\usepackage{xcolor}
\usepackage{tcolorbox}
\tcbuselibrary{listings,breakable,skins}

\newtcblisting{promptbox}[2][]{%
  enhanced,
  breakable,
  colback=black!2,
  colframe=black!25,
  boxrule=0.35pt,
  arc=1.2mm,
  left=2mm,right=2mm,top=1mm,bottom=1mm,
  fonttitle=\bfseries,
  title={#2},
  listing only,
  listing options={
    basicstyle=\ttfamily\footnotesize,
    columns=fullflexible,
    breaklines=true,
    breakatwhitespace=false,
    showstringspaces=false,
    tabsize=2
  },
  #1
}

\usepackage[preprint]{acl}

\usepackage{times}
\usepackage{latexsym}

\usepackage[T1]{fontenc}

\usepackage[utf8]{inputenc}

\usepackage{microtype}

\usepackage{inconsolata}

\usepackage{graphicx}

\title{S2G-RAG: Structured Sufficiency and Gap Judging for Iterative Retrieval-Augmented QA}

\author{
Minghan Li\thanks{Equal contribution.}\thanks{Corresponding author.},
Junjie Zou\footnotemark[1],
Xinxuan Lv,
Chao Zhang,
Guodong Zhou \\
Soochow University, Suzhou, China \\
\texttt{mhli@suda.edu.cn, jjzou1@stu.suda.edu.cn, xxlv@stu.suda.edu.cn,} \\
\texttt{czhang1@stu.suda.edu.cn, gdzhou@suda.edu.cn}
}

\begin{document}
\maketitle
\begin{abstract}
Retrieval-Augmented Generation (RAG) grounds language models in external evidence, but multi-hop question answering remains difficult because iterative pipelines must control what to retrieve next and when the available evidence is adequate. In practice, systems may answer from incomplete evidence chains, or they may accumulate redundant or distractor-heavy text that interferes with later retrieval and reasoning.
We propose \textbf{S2G-RAG} (\textbf{S}tructured \textbf{S}ufficiency and \textbf{G}ap-judging RAG), an iterative framework with an explicit controller, S2G-Judge. At each turn, S2G-Judge predicts whether the current evidence memory supports answering and, if not, outputs structured gap items that describe the missing information. We map these gap items into the next retrieval query, producing stable multi-turn retrieval trajectories. To reduce noise accumulation, we maintain a sentence-level Evidence Context by extracting a compact set of relevant sentences from retrieved documents.
Experiments on TriviaQA, HotpotQA, and 2WikiMultiHopQA show that S2G-RAG improves multi-hop QA performance and robustness under multi-turn retrieval. Furthermore, S2G-RAG can be integrated into existing RAG pipelines with a lightweight component, without modifying the search engine or retraining the generator.

\end{abstract}

\section{Introduction}
\label{sec:intro}

Retrieval-augmented generation (RAG) grounds large language models (LLMs) in external evidence and has become a standard approach for knowledge-intensive question answering~\citep{lewis2020retrieval, shuster2021retrieval, arslan2024survey}.
While single-hop questions can often be answered from one document, multi-hop question answering requires composing evidence across multiple documents, where later retrieval depends on intermediate findings from earlier rounds~\citep{yang2018hotpotqa, tang2024multihop}.
This motivates iterative RAG, which interleaves retrieval and reasoning over multiple turns.
However, iterative RAG faces a retrieval control bottleneck.
At each turn, the system must assess whether the accumulated evidence supports answering, and if not, specify what to retrieve next~\citep{yang2025knowing, ye2025qprm}.
When control is unreliable, systems either answer from incomplete evidence chains or continue retrieving without a clear target, accumulating redundant or distractor-heavy text that interferes with subsequent retrieval and reasoning.

A Hotpot-style question such as ``In which country was the director of \textit{Home Alone} born?'' illustrates the challenge.
An initial retrieval may identify the bridge entity (Chris Columbus) but still miss the answer-bearing attribute (his country of birth).
Answering at this point fails, whereas continuing with generic follow-up queries can repeatedly retrieve near-duplicate pages and grow noisy contexts.
Effective iterative RAG therefore requires an auditable turn-level state that reflects whether the current evidence is adequate and, when it is not, what concrete information gap remains.

Recent work improves iterative retrieval via query rewriting, decomposition, intermediate reasoning traces, and self-reflection or critic signals~\citep{trivedi2023interleaving, asai2024self, yang2025knowing, dong2025rag, wang2025llms, jiang2023active}.
Despite progress, three limitations remain.
First, control is often implicit and entangled with free-form generation, which makes decisions difficult to audit and brittle under distractors.
Second, the representation of the next-hop information need is under-specified, leading to drift, repetition, or over-generalization when intermediate contexts are noisy.
Third, robust control benefits from turn-level supervision, but intermediate-state annotation is expensive~\citep{yang2025knowing}, and supervision from idealized trajectories can misalign with the intermediate evidence contexts produced by multi-turn pipelines.
Meanwhile, iterative retrieval naturally accumulates long and noisy text, and naive concatenation can further destabilize control~\citep{hwang2025exit, jiang2025retrieve, liu2024lost}.

We propose \textbf{S2G-RAG} (\textbf{S}tructured \textbf{S}ufficiency and \textbf{G}ap-judging RAG), an iterative framework that makes retrieval control explicit and modular.
Its lightweight controller, S2G-Judge, takes the question $q$ and the accumulated evidence context $C_t$ and outputs a binary sufficiency decision together with structured gap items that describe the missing information for the next hop.
S2G-RAG maps the predicted gap items into the next retrieval query, making the next-hop target explicit and diagnosable.
The controller is decoupled from answer generation, so the answer reasoner operates only on the accumulated evidence context.

To mitigate noise accumulation, S2G-RAG maintains a compact sentence-level evidence context.
Given newly retrieved documents, an LLM-based Evidence Extractor selects salient sentences and explicitly prioritizes those aligned with the predicted gap items, producing an auditable evidence context for subsequent judging and final answering.

We train S2G-Judge with process supervision distilled from multi-turn pipeline execution traces.
We run the retrieval and evidence accumulation procedure on training questions and log per-turn snapshots $(q, C_t)$ by rolling out to a fixed maximum budget.
A strong teacher then labels each snapshot under a context-only evidence constraint with sufficiency decisions and gap items, and we distill these signals into a compact judge model~\citep{lightman2023let, hsieh2023distilling}.

Our main contributions are as follows.
\begin{itemize}
    \item \textbf{Structured sufficiency and gap judging.} We formulate iterative RAG control as per-turn structured prediction of evidence sufficiency and explicit information gaps, implemented via a lightweight judge decoupled from answer generation.
    \item \textbf{Distillation from execution traces.} We introduce a supervision scheme built from multi-turn execution traces, enabling turn-level learning under realistic intermediate evidence contexts.
    \item \textbf{Gap-aware sentence-level evidence context.} We maintain a compact, auditable evidence context via pointer-based extraction with explicit gap-aware prioritization to reduce distractor interference.
    \item \textbf{Empirical validation.} Across TriviaQA, HotpotQA, and 2WikiMultiHopQA, S2G-RAG improves EM/F1 over strong baselines under both sparse and dense retrieval settings.
\end{itemize}

\section{Related Work}

\subsection{Retrieval-Augmented Generation}
Retrieval-augmented generation (RAG) grounds LLM outputs in external corpora by retrieving relevant documents and conditioning generation on the retrieved evidence~\citep{lewis2020retrieval, shuster2021retrieval, arslan2024survey}.
Most standard RAG systems follow a single-round retrieve-then-generate pattern, which is effective when the required evidence can be covered in one retrieval pass.
However, many knowledge-intensive questions require multi-step reasoning over multiple sources, including synthesizing evidence across documents, aligning entities and relations, and resolving information needs that only become clear after inspecting initial retrieval results.
This motivates multi-round retrieval-and-reasoning systems that iteratively update an information state before producing the final answer.

\subsection{Multi-Round RAG}
Multi-hop question answering requires integrating multiple evidence pieces and performing multi-step reasoning, and is widely used to evaluate both RAG systems and LLM reasoning capabilities.
In this setting, a single retrieval pass is often insufficient because crucial evidence is distributed across documents and later-hop retrieval depends on intermediate findings from earlier rounds.
Recent work therefore proposes iterative RAG frameworks that alternate between retrieval and reasoning over multiple rounds.

We organize related work along four themes that recur in iterative retrieval control and evidence management.
These themes include intermediate query generation, stopping criteria, supervision for turn-level control, and context management under multi-turn noise and bloat.

\paragraph{Intermediate query generation.}
A major line of work improves multi-round retrieval by generating intermediate queries, including explicit planning and question decomposition into sub-questions, as well as follow-up query rewriting from partial evidence~\citep{ye2025qprm, jiang2023active, dong2025rag, wang2025llms}.
While these approaches can improve later-hop recall and evidence composition, the intermediate information need is often expressed in free-form text.
As a result, query generation can drift, repeat, or become unstable under distractors and noisy intermediate contexts, and reliable inference-time behavior may require repeated calls to strong models.

\paragraph{Stopping criteria.}
Another line of work studies when it is safe to answer, which is commonly framed as deciding whether the current evidence is sufficient~\citep{yang2025knowing}.
Several methods rely on self-reflection or critic-style verification to decide whether to continue retrieval~\citep{asai2024self, yang2025knowing}.
These signals can reduce incorrect answers produced under incomplete evidence, but the control state is often not standardized or easily auditable.
Moreover, when retrieval continues, these approaches typically provide limited explicit structure for specifying the remaining information need.

\paragraph{Supervision for turn-level control.}
Learning reliable turn-level control benefits from process-level supervision, yet scalable, high-quality labels for intermediate decisions and next-step information needs are difficult to obtain~\citep{yang2025knowing}.
Prior work explores distilling supervision from stronger teachers or leveraging trajectory-based learning signals for multi-step decision making~\citep{lightman2023let, hsieh2023distilling}.
A recurring challenge is distribution mismatch, where supervision derived from synthetic trajectories or heuristic signals may not reflect the intermediate evidence states produced by multi-turn pipelines in practice, which often include redundancy and distractors.

\paragraph{Context management.}
Iterative retrieval tends to accumulate long and noisy contexts, which weakens evidence focus and can destabilize both control and reasoning~\citep{liu2024lost}.
To mitigate context bloat, prior work compresses retrieved content or performs sentence-level evidence selection~\citep{dhole2025retrieve, hwang2025exit}.
These techniques are complementary to retrieval control because they reduce the influence of redundant or distractor-heavy text, but they do not by themselves specify what to retrieve next or provide an explicit representation of the missing information that drives subsequent retrieval.
\begin{figure*}[t]
  \centering
  \includegraphics[width=\textwidth]{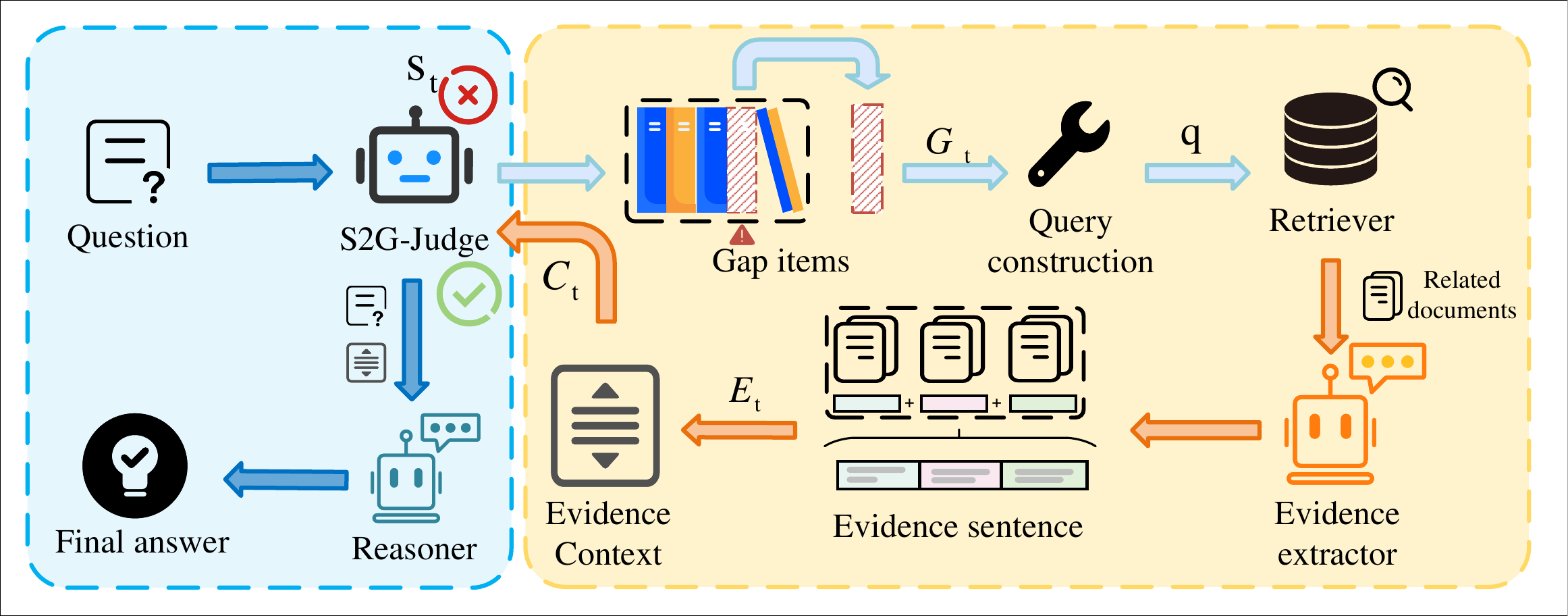}
  \caption{Overview of the S2G-RAG inference framework. Given a question $q$ and the current evidence context $C_t$, S2G-Judge first predicts a sufficiency decision $s_t$ and structured gap items $G_t$. If $s_t=\texttt{true}$, a reasoner generates the final answer from $(q, C_t)$; otherwise, $G_t$ guides query construction for retrieval, and a sentence-level evidence extractor selects salient evidence sentences from retrieved documents to form $E_t$, which is appended to update $C_t=C_{t-1}\oplus E_t$. The process iterates until the evidence is sufficient or the retrieval budget is reached.}
  \label{fig:s2g_rag_framework}
\end{figure*}

\section{Methodology}
\label{sec:method}

This section presents \textbf{S2G-RAG}, an iterative RAG framework for multi-hop QA that makes retrieval control explicit through structured sufficiency and gap judging.
We first describe the S2G-RAG inference loop (Section~\ref{sec:s2g_rag_framework}), then introduce S2G-Judge training via trajectory distillation from execution traces (Section~\ref{sec:s2g_judge}), and finally detail gap-guided query construction and sentence-level evidence extraction (Sections~\ref{sec:query_construction}--\ref{sec:evidence_extractor}).

\subsection{S2G-RAG Inference Framework}
\label{sec:s2g_rag_framework}

Given a question $q$ and an external corpus $\mathcal{D}$, S2G-RAG maintains an accumulated evidence context $C_t$ at turn $t$.
We initialize $C_0$ as empty and run the loop for at most $T$ retrieval turns.
Figure~\ref{fig:s2g_rag_framework} illustrates the overall framework.
At each turn, S2G-RAG follows a judge-first loop with four components: S2G-Judge, a retriever, a sentence-level evidence extractor, and an answer reasoner.

\paragraph{S2G-Judge.}
At turn $t$, the judge reads only $(q, C_t)$ and outputs $y_t=(s_t, G_t)$.
The binary variable $s_t\in\{\texttt{true},\texttt{false}\}$ indicates whether the current evidence context is sufficient to support answering, and $G_t$ is a set of structured \emph{gap items} (Section~\ref{sec:judge_schema}).
When $s_t=\texttt{true}$, the system proceeds to answer using $(q, C_t)$.
This sufficiency signal is primarily used to reduce answers produced under incomplete evidence.

\paragraph{Retriever.}
When $s_t=\texttt{false}$, the system constructs a query $\tilde{q}_t$ from $G_t$ (Section~\ref{sec:query_construction}) and retrieves top-$k$ documents $D_t=\{d_{t,1},\ldots,d_{t,k}\}$ from $\mathcal{D}$.

\paragraph{Evidence extractor.}
A sentence-level extractor $h$ selects salient sentences from $D_t$ conditioned on $(q, G_t)$ and forms a verbatim evidence block $E_t$.
We update the evidence context by concatenation
\begin{equation}
C_t = C_{t-1} \oplus E_t,
\end{equation}
where $\oplus$ denotes concatenation.

\paragraph{Reasoner.}
The answer reasoner $r_\theta$ is decoupled from retrieval control.
It is invoked only when the system decides to answer or when the loop reaches the maximum budget.
In either case, the reasoner produces the final answer conditioned on $(q, C_t)$.

A pseudocode description of the inference loop is provided in Appendix~\ref{sec:pseudocode}.

\subsection{S2G-Judge}
\label{sec:s2g_judge}

\subsubsection{Output Schema and Constraints}
\label{sec:judge_schema}

At each turn $t$, S2G-Judge predicts a structured output
\begin{equation}
y_t = (s_t, G_t), \quad s_t \in \{\texttt{true},\texttt{false}\}.
\end{equation}
If $s_t=\texttt{true}$, the judge outputs an empty gap set, so $G_t=\emptyset$.

Each gap item $g \in G_t$ follows a normalized schema with four fields.
The \texttt{category} field takes values in \{\texttt{bridge\_entity}, \texttt{attribute}, \texttt{relation}, \texttt{evidence\_span}, \texttt{other}\}.
The \texttt{target} field specifies the entity being queried.
The \texttt{slot} field indicates a coarse-grained attribute or relation name.
The \texttt{description} field provides a short natural-language clarification.

We impose a context-only constraint for sufficiency judgments.
The judge must determine sufficiency based strictly on evidence present in $C_t$, which prevents relying on parametric knowledge when deciding whether the evidence is adequate.

\subsubsection{Training via Trajectory Distillation from Execution Traces}
\label{sec:judge_data}

We train S2G-Judge with process supervision distilled from multi-turn execution traces.
We execute the same iterative retrieval and evidence accumulation procedure on training questions and log per-turn states $(q, C_t)$.
During trace collection, we instantiate the control loop with an unfine-tuned judge backbone under the same prompt/interface, and roll out the retrieval process to the maximum budget $T$.
This produces intermediate evidence contexts that reflect multi-turn accumulation, including redundancy and distractors, which serve as inputs for judge training.

A stronger teacher model labels each $(q, C_t)$ under the context-only constraint.
We apply lightweight conflict filtering to remove low-confidence supervision. We then perform LoRA-based supervised fine-tuning so that the judge directly predicts the structured output $y_t$ conditioned on $x_t=(q, C_t)$
\begin{equation}
\mathcal{L}(\phi) = - \sum_{(x_t,y_t)} \sum_{i=1}^{|y_t|}
\log p_{\phi}\big(y_{t,i} \mid y_{t,<i}, x_t\big).
\end{equation}

Additional training details, including supervision construction and hyperparameters, are provided in Appendix~\ref{sec:training_details}.

\subsection{Gap-Guided Query Construction}
\label{sec:query_construction}

When $s_t=\texttt{false}$, S2G-RAG constructs the next retrieval query $\tilde{q}_t$ by mapping predicted gap items to query phrases.
For each gap item, we prioritize concatenating \texttt{target} and \texttt{slot} when both fields are present, and otherwise fall back to the \texttt{description} field.
We take up to the first $K$ valid phrases in the order produced by the judge and append them to the original question.
If no valid phrase can be constructed, we set $\tilde{q}_t=q$.

\subsection{Sentence-level Evidence Extractor}
\label{sec:evidence_extractor}

LLMs are sensitive to noisy or misleading evidence~\citep{shi2023large, zeng2025worse} and can degrade under long inputs~\citep{liu2024lost}.
In iterative retrieval, directly concatenating full documents across turns can rapidly expand $C_t$ and amplify distractors.
We therefore introduce a sentence-level evidence extractor $h$ that converts newly retrieved documents into a compact evidence block before updating the accumulated evidence context.

At turn $t$, the retriever returns documents $D_t=\{d_{t,1},\ldots,d_{t,k}\}$.
We split each document into sentences and form a globally indexed candidate pool $\mathcal{S}_t$, where each candidate is paired with its document title for provenance.
The extractor is an LLM prompted with $(q, G_t, \mathcal{S}_t)$ and is constrained to output only indices from $\mathcal{S}_t$, without rewriting or generating any evidence text.
We then deterministically map the selected indices back to the original sentences and concatenate them to form the evidence block:
\begin{equation}
E_t = h(q, G_t, D_t), \qquad C_t = C_{t-1} \oplus E_t.
\end{equation}

The extractor is designed to preserve multi-hop progress under noisy retrieval by selecting answer-bearing sentences, bridge facts, and disambiguating context for entities in $q$. The predicted gap items $G_t$ provide an explicit signal for prioritizing evidence that addresses the current missing information need. This sentence-level evidence context limits redundant and distractor-heavy text in top-$k$ retrieval results, helping keep downstream judging and reasoning focused, while its pointer-based design preserves auditability and reduces the risk of hallucinated evidence.

\begin{table*}[t]
\centering
\small
\setlength{\tabcolsep}{4pt}
\begin{tabular}{l l cc cc cc}
\toprule
Method & Reasoner & \multicolumn{2}{c}{TriviaQA} & \multicolumn{2}{c}{HotpotQA} & \multicolumn{2}{c}{2Wiki} \\
& / Retriever & EM & F1 & EM & F1 & EM & F1 \\
\midrule

\multicolumn{8}{c}{\textit{Panel A: Sparse retrieval (BM25)}} \\
NaiveGen        & Llama3-8B / --          & 55.7 & 63.1 & 20.6 & 28.4 & 26.4 & 33.9 \\
IR-CoT          & Llama3-8B / BM25        & 56.9 & 68.9 & 28.6 & 41.5 & 23.5 & 32.4 \\
SIM-RAG         & Llama3-8B / BM25        & 70.7 & 75.6 & 32.7 & 43.3 & 34.1 & 40.2 \\
\midrule
S2G-RAG (ours)  & Llama3-8B / BM25        & \textbf{72.0} & \textbf{77.9} & \textbf{43.3} & \textbf{56.5} & \textbf{41.7} & \textbf{48.6} \\
\midrule

\multicolumn{8}{c}{\textit{Panel B: Dense retrieval (E5)}} \\
Standard RAG    & Llama3-8B / E5          & 58.8 & 68.3 & 25.1 & 35.3 & 10.6 & 21.0 \\
Self-RAG        & Llama3-8B / E5          & 38.2 & 53.4 & 17.1 & 29.6 & 12.1 & 25.1 \\
FLARE           & Llama3-8B / E5        & 55.8 & 63.2 & 19.7 & 28.0 & 25.8 & 33.9 \\
ReSP$^{\star}$  & Llama3-8B / E5          & --   & --   & --   & 47.2 & --   & 38.3 \\
RAG-Critic      & Llama3-8B / E5        & 65.0 & 75.9 & 40.0 & 51.2 & 27.9 & 34.0 \\
\midrule
S2G-RAG (ours)  & Llama3-8B / E5          & \textbf{71.1} & \textbf{78.0} & \textbf{42.0} & \textbf{53.5} & \textbf{39.0} & \textbf{45.3} \\
\bottomrule
\end{tabular}
\caption{Main results (EM/F1) on TriviaQA, HotpotQA, and 2WikiMultiHopQA.
Panel A reports methods under BM25 and Panel B reports methods under E5.
Bold denotes the best score within each panel under the same retriever setting.
ReSP$^{\star}$ results are reported from the original paper for reference.}
\label{tab:main_results}
\end{table*}

\section{Experiments}

\subsection{Task and Datasets}
We evaluate \textbf{S2G-RAG} on three open-domain QA benchmarks spanning single-hop and multi-hop reasoning.
For single-hop QA, we use TriviaQA~\citep{joshi2017triviaqa}, where answers are often supported by a single Wikipedia article.
For multi-hop QA, we adopt HotpotQA~\citep{yang2018hotpotqa}, which requires composing evidence across multiple documents, and 2WikiMultiHopQA~\citep{ho2020constructing}, which emphasizes entity disambiguation and fine-grained evidence composition.
Following standard evaluation protocols, we report Exact Match (EM) and F1, and use the Wikimedia dumps provided by each dataset as the retrieval corpus.

\subsection{Implementation Details}
\label{sec:impl_details}
We instantiate the answer reasoner with Llama-3-8B-Instruct.
For simplicity, the Evidence Extractor uses the same backbone as the reasoner.

For retrieval control, we fine-tune a lightweight backbone with LoRA to obtain S2G-Judge variants based on Llama-3.2-3B-Instruct.
Training supervision is distilled from multi-turn execution traces labeled by a stronger teacher model (GPT-4o-mini), following Section~\ref{sec:judge_data}.
Unless otherwise noted, S2G-RAG runs for at most $T{=}4$ retrieval turns, retrieves top-$k{=}6$ documents per turn, and applies title-based de-duplication across turns.

We report results under both sparse and dense retrieval settings.
For sparse retrieval, we use BM25 implemented in Pyserini.
For dense retrieval, we use E5-base-v2~\citep{wang2022text}.
All methods share the same retrieval corpus for each dataset and are evaluated on the official development split.
Our pipeline can be replicated with 2$\times$vGPU-48G or equivalent hardware.
LoRA configurations, decoding settings and additional training details are included in the Appendix~\ref{sec:training_details}.

\subsection{Baselines}
\label{sec:baselines}
We compare S2G-RAG against representative baselines spanning single-round RAG and multi-round retrieval control.
NaiveGen~\citep{jin2025flashrag} answers without retrieval.
Standard RAG~\citep{jin2025flashrag} performs a single retrieval using the original question and generates an answer conditioned on the retrieved documents.
IR-CoT~\citep{trivedi2023interleaving} interleaves retrieval with intermediate reasoning signals that serve as retrieval cues.
FLARE~\citep{jiang2023active} triggers retrieval based on generation-time uncertainty signals.
ReSP~\citep{jiang2025retrieve} iterates retrieve--summarize--plan to manage multi-turn evidence accumulation.
Self-RAG~\citep{asai2024self} fine-tunes a generator to emit explicit reflection tokens that govern retrieval behavior.
SIM-RAG~\citep{yang2025knowing} trains a lightweight critic that, at each round, evaluates the reasoner’s provisional answer and accompanying rationale against the currently retrieved context and decides whether to accept the answer or continue searching. RAG-Critic~\citep{dong2025rag} introduces an error-aware critic that is trained with a hierarchical error taxonomy and provides fine-grained feedback about potential RAG failures, which is then used to trigger targeted correction workflows in an agentic RAG pipeline.

\subsection{Main Results}
\label{sec:main_results}

Table~\ref{tab:main_results} reports end-to-end EM/F1 on TriviaQA, HotpotQA, and 2WikiMultiHopQA under both sparse (BM25) and dense (E5) retrieval.
Across settings, S2G-RAG improves over strong baselines, supporting the benefit of making retrieval control explicit through structured sufficiency and gap judging.

Under a matched BM25 setting with the same answer reasoner, S2G-RAG substantially outperforms prior multi-round baselines.
Relative to SIM-RAG, S2G-RAG improves TriviaQA by $+1.3$ EM and $+2.3$ F1 (from 70.7/75.6 to 72.0/77.9).
On HotpotQA, the gains are larger at $+10.6$ EM and $+13.2$ F1 (from 32.7/43.3 to 43.3/56.5), consistent with the stronger need for accurate multi-hop progress.
On 2WikiMultiHopQA, S2G-RAG yields a $+7.6$ EM and $+8.4$ F1 improvement (from 34.1/40.2 to 41.7/48.6).
The larger gains on the multi-hop benchmarks suggest that structured gap prediction and controlled evidence accumulation are especially helpful when later retrieval depends on intermediate evidence.

S2G-RAG also transfers across retriever families.
With E5, S2G-RAG attains the best results within the dense-retrieval panel on all three benchmarks, outperforming Standard RAG and learned-control baselines reported in this setting such as RAG-Critic.
This suggests that the proposed control interface is not tied to a particular retriever.

Finally, the TriviaQA results highlight a common challenge for iterative pipelines.
When only a small amount of evidence is needed, additional retrieval can introduce redundant or distracting text that interferes with answer generation.
S2G-RAG mitigates this issue by maintaining a compact sentence-level evidence context, so that increased retrieval does not directly translate into uncontrolled context growth.
On multi-hop benchmarks, the improvements are larger because explicit gap specification and controlled evidence accumulation better support multi-step evidence composition.

\subsection{Analysis Experiments}

\subsubsection{Ablation Study}
\label{sec:ablation}

Table~\ref{tab:ablation} reports ablations on HotpotQA (dev) under BM25. Starting from the full pipeline, we remove one component at a time while keeping the retriever, retrieval budget, and answer reasoner fixed.

Removing S2G-Judge causes the largest drop ($-15.8$ EM, $-19.0$ F1), confirming the importance of explicit retrieval control for multi-hop QA. Without the judge, the system loses both the sufficiency check and the structured gap signals that guide the next retrieval step.

Replacing the trained judge with its unfine-tuned backbone also hurts performance ($-4.1$ EM, $-5.7$ F1), showing the value of trajectory distillation on intermediate evidence contexts. Notably, even without supervised fine-tuning, this variant still outperforms prior BM25 baselines on HotpotQA, suggesting that the structured control interface is useful even before training.

Removing the sentence-level Evidence Extractor yields a further drop ($-3.8$ EM, $-4.0$ F1), highlighting its role in filtering redundant or distractor-heavy retrieved text and keeping judging and answering focused on salient evidence.

\begin{table}[t]
\centering
\small
\setlength{\tabcolsep}{5pt}
\begin{tabular}{l cc}
\toprule
Variant & EM & F1 \\
\midrule
Full S2G-RAG                 & \textbf{43.3} & \textbf{56.5} \\
w/o SFT (Untrained Judge)    & 39.2 ($-4.1$)  & 50.8 ($-5.7$) \\
w/o S2G-Judge                & 27.5 ($-15.8$) & 37.6 ($-19.0$) \\
w/o Extractor                & 39.5 ($-3.8$)  & 52.5 ($-4.0$) \\
\bottomrule
\end{tabular}
\caption{Ablation on HotpotQA (dev) under BM25. Starting from the full S2G-RAG pipeline, we remove one component at a time while keeping the retriever, maximum turns $T$, per-turn top-$k$, and the answer reasoner fixed. w/o SFT replaces the LoRA-finetuned S2G-Judge with its unfine-tuned backbone under the same prompt and output schema. w/o S2G-Judge removes the judge module, so the pipeline runs without structured sufficiency or gap predictions. w/o Extractor appends raw retrieved text instead of sentence-level evidence blocks. Parentheses indicate absolute drops from the full system.}
\label{tab:ablation}
\end{table}

\subsubsection{Analysis on Sufficiency Predictions} \label{sec:sufficiency_analysis} To assess the reliability of the judge’s sufficiency signal, we report the confusion matrix of S2G-Judge on HotpotQA (dev) for its binary sufficient/insufficient decisions.
\begin{figure}[t]
  \centering
  \includegraphics[width=\columnwidth]{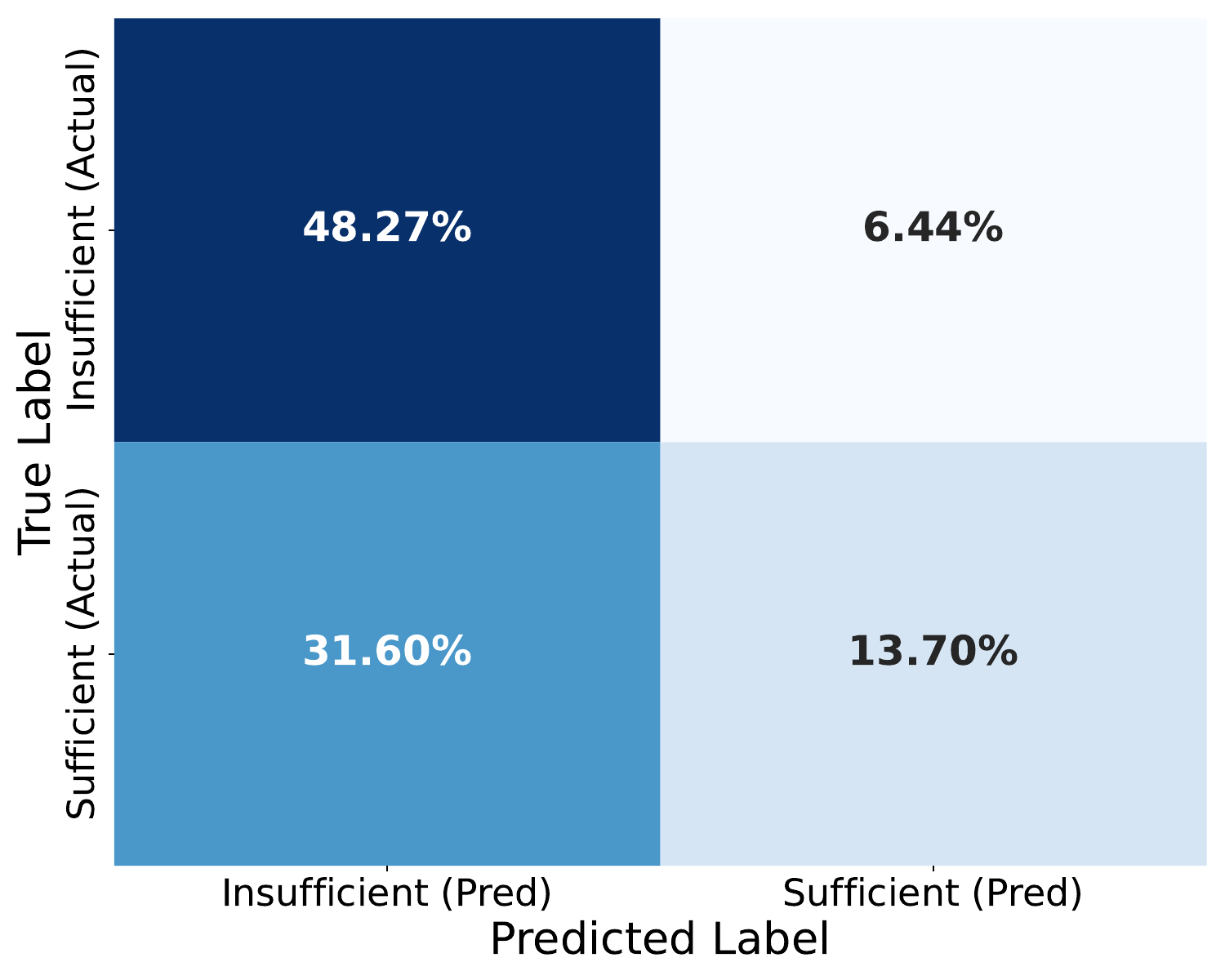}
  \caption{Confusion matrix of S2G-Judge's binary sufficiency decisions on HotpotQA (dev) using BM25 retrieval, with GPT-4o-mini as the reasoner and the evidence extractor. Here we use a retrieval-only ground-truth label $\textsc{Truth}=\textsc{CorrectRetrieval}$, where \textsc{CorrectRetrieval} is true iff the accumulated retrieved \emph{titles} cover all gold supporting document titles for the instance. Values are percentages over the evaluated set.}
  \label{fig:suff_cm}
\end{figure}
Figure~\ref{fig:suff_cm} indicates that S2G-Judge is unlikely to over-claim sufficiency. Conditioned on predicting sufficient, the judge is correct under Truth in most cases. In particular, the false-positive rate is 6.44\%. This behavior is aligned with the role of the sufficiency signal in S2G-RAG, namely reducing answers produced from evidence contexts that are not yet adequate. Meanwhile, the judge tends to require more explicit support in the accumulated context before it declares sufficient. A portion of examples that satisfy Truth are still predicted as insufficient (31.60\%). This pattern suggests that the judge is more prone to under-claim sufficiency than to prematurely accept borderline evidence contexts, leaving room for improving calibration so that clearly sufficient contexts are more often recognized as such.

\subsubsection{Efficiency of Evidence Memory Compression}
\label{sec:compression}

\begin{figure}[t]
  \centering
  \includegraphics[width=\columnwidth]{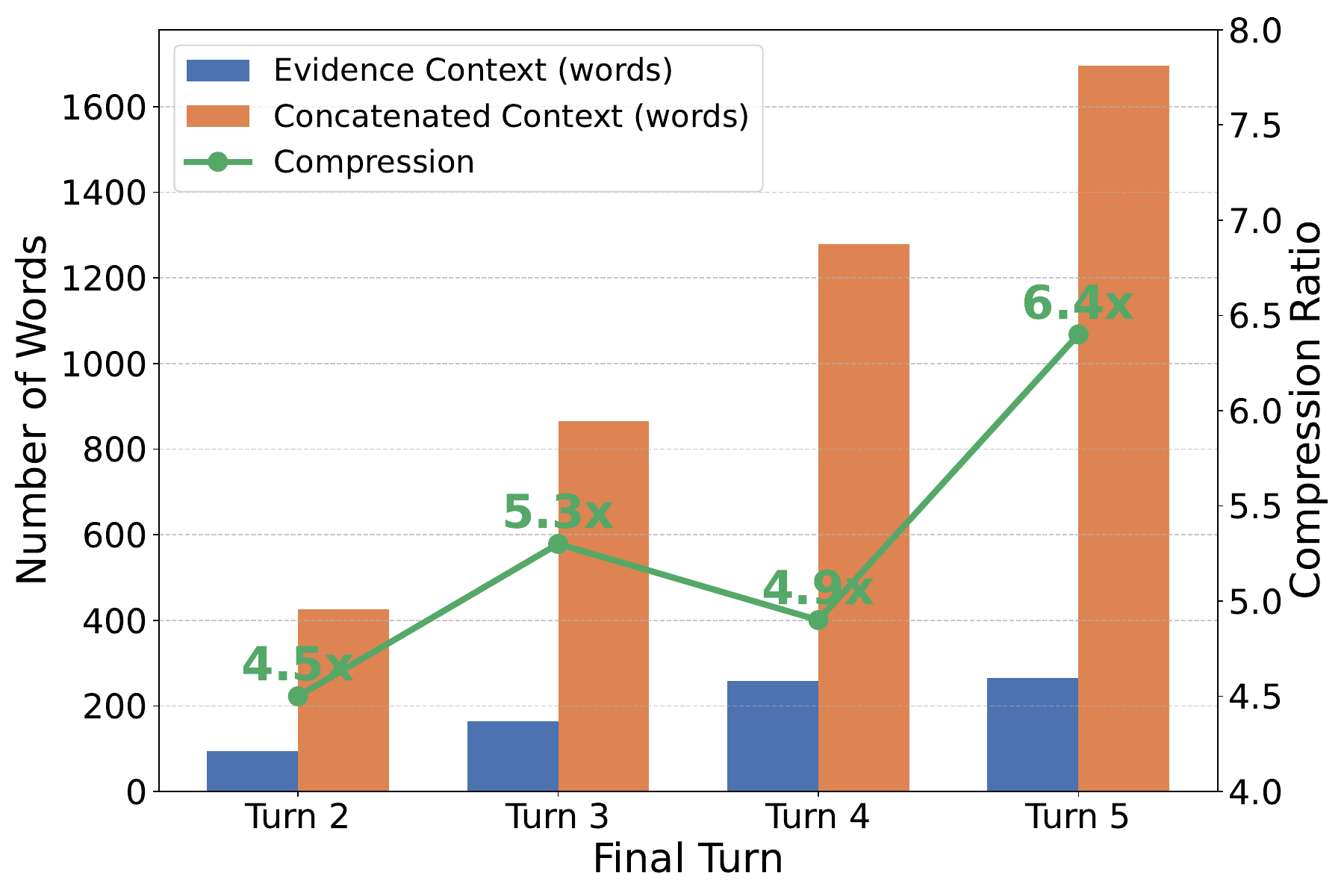}
  \caption{Context compression on HotpotQA (dev), computed on the final row per question ID.
Evidence Context accumulates only extractor-selected sentence blocks, while Concatenated Context accumulates raw retrieved documents across turns.
We measure context length by whitespace-separated word counts.
Bars show mean context length grouped by the final turn; $N$ denotes the number of questions whose trajectories end at that turn.
The line reports the compression ratio (Concatenated/Evidence).}
  \label{fig:compression_by_turn}
\end{figure}

Figure~\ref{fig:compression_by_turn} shows that the sentence-level Evidence Extractor effectively controls context growth over multi-turn retrieval. Compared with raw document concatenation, the resulting Evidence Context remains substantially more compact across all final-turn groups, yielding a compression ratio of roughly $4.5\times$--$6.4\times$. The advantage becomes more pronounced for examples that require more retrieval turns, where raw context accumulation introduces increasing redundancy and distractors.

\begin{table}[t]
\centering
\small
\setlength{\tabcolsep}{4pt}
\begin{tabular}{lccc}
\toprule
Method & EM & F1 & Per-sample Latency (s/q) \\
\midrule
Standard RAG & 25.1 & 35.3 & 0.3787 \\
Ours w/o extractor & 39.5 & 52.5 & 1.9552 \\
S2G-RAG (ours) & \textbf{43.3} & \textbf{56.5} & 1.6085 \\
\bottomrule
\end{tabular}
\caption{Latency evaluation on HotpotQA under BM25 in the same environment as the main results. Standard RAG uses one retrieval round, while the iterative variants use at most $T \leq 4$ retrieval turns.}
\label{tab:latency}
\end{table}

Table~\ref{tab:latency} further shows that evidence memory compression improves runtime efficiency in multi-turn retrieval. Although iterative retrieval is slower than single-round RAG, it brings large gains in QA accuracy. More importantly, compared with appending full retrieved text at each turn, the sentence-level extractor reduces per-sample latency from 1.9552 to 1.6085 seconds while also improving EM/F1, corresponding to a 17.7\% latency reduction together with gains of +3.8 EM and +4.0 F1. These results suggest that controlling evidence memory improves not only context compactness but also the efficiency of multi-turn retrieval.

\subsubsection{Comparison with Evidence Compression Baselines}

To study the trade-off between context compactness and downstream QA accuracy, we compare our sentence-pointer extractor with LLM summarization and ReComp \citep{xu2024recomp} under the same HotpotQA pipeline setting.

\begin{table}[t]
\centering
\small
\begin{tabular}{lccc}
\toprule
Method & EM & F1 & Compression ratio \\
\midrule
LLM summarization & 36.9 & 48.1 & 0.3816 \\
ReComp (extractive) & 41.4 & 53.8 & 0.4948 \\
ReComp (abstractive) & 34.9 & 46.4 & 0.1917 \\
Ours (sentence pointers) & \textbf{43.3} & \textbf{56.5} & 0.3461 \\
\bottomrule
\end{tabular}
\caption{Comparison of evidence compression methods on HotpotQA under the same iterative QA pipeline setting. All variants use the same retriever, retrieval budget, and answer reasoner, and differ only in the evidence compression module. Compression ratio is measured as compressed-context words divided by raw retrieved-context words.}
\label{tab:compression_baselines}
\end{table}

As shown in Table~\ref{tab:compression_baselines}, abstractive compression produces shorter contexts but leads to a clear drop in EM/F1. ReComp improves over generic summarization, yet both its extractive and abstractive variants remain below our method. Our extractor achieves the best EM/F1 while still maintaining substantial compression with a compression ratio of 0.3461, indicating a better trade-off between compression and QA accuracy.

\subsubsection{Teacher--Student Gap under Varying Retrieval Budgets}
\label{sec:teacher_student_gap}
\begin{figure}[t]
  \centering
  \includegraphics[width=\columnwidth]{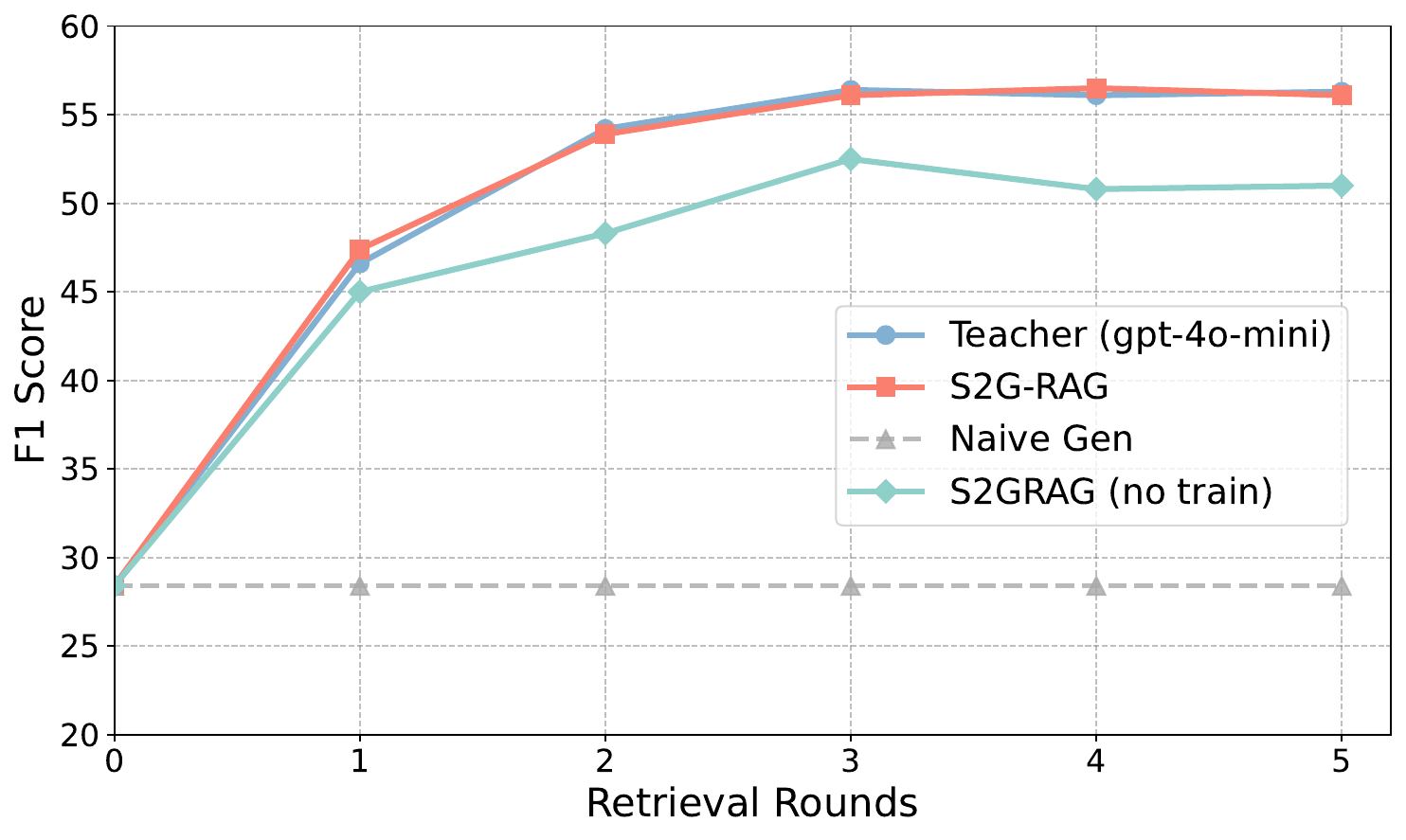}
\caption{F1 on HotpotQA (dev) under varying maximum retrieval budget $T$. We compare a teacher controller (GPT-4o-mini), S2G-RAG with the trained S2G-Judge, an unfine-tuned controller (no train), and NaiveGen without retrieval. In the teacher setting, GPT-4o-mini replaces only the S2G-Judge/controller; all other components and retriever settings are unchanged. It receives only the question and accumulated evidence context, and the retrieval query is deterministically constructed by our system from the predicted gap items. All variants use the same answer reasoner and sentence-level evidence extractor with the same maximum sentence cap.}
  \label{fig:f1_vs_budget}
\end{figure}
To study the teacher--student gap under different retrieval budgets, we vary the maximum number of retrieval turns $T$ from 0 to 5 while keeping all other components fixed. Since all variants share the same answer reasoner and evidence extractor, the differences mainly reflect controller quality.

Figure~\ref{fig:f1_vs_budget} shows large gains when increasing $T$ from small budgets, followed by saturation at larger budgets. Across the full sweep, the trained S2G-Judge closely tracks the teacher controller, suggesting that trajectory distillation transfers much of the teacher's control behavior to a smaller model.

The saturation at larger budgets is consistent with multi-hop retrieval. Once the key supporting evidence has been accumulated, additional turns are less likely to contribute new answer-relevant information. The sentence-level extractor further limits context growth by retaining only salient evidence, so later retrieval results are increasingly filtered as redundant or weakly related.

In contrast, the unfine-tuned controller performs consistently worse and scales less reliably as $T$ increases, highlighting the importance of distillation for stable sufficiency and gap prediction. NaiveGen remains flat across budgets, confirming that the gains come from retrieval and controller behavior rather than additional generation alone.

\section{Conclusion}
\label{sec:conclusion}
In this paper, we propose S2G-RAG, which turns two key control decisions into explicit, structured outputs.
At each turn, a lightweight S2G-Judge assesses whether the current evidence context is adequate and, when it is not, produces structured gap items that describe what information is still needed.
These gap items provide a concrete handle for constructing the next retrieval query, reducing reliance on free-form query rewriting and making retrieval trajectories easier to inspect.
To keep multi-turn contexts reliable under distractor-heavy retrieval, we complement this control interface with sentence-level evidence extraction that retains a compact, provenance-preserving Evidence Context for subsequent turns and final answering.

Empirically, S2G-RAG improves EM/F1 on TriviaQA, HotpotQA, and 2WikiMultiHopQA under both BM25 and E5 retrieval.
Our analysis suggests the judge reduces answers made under insufficient evidence, while sentence-level extraction limits the influence of redundant or distracting text.
We hope these findings motivate further work on structured and auditable control mechanisms for iterative RAG systems.

\section*{Limitations}
\label{sec:limitations}

We acknowledge several limitations and opportunities for future work.
First, recent work has explored learning-based optimization for iterative retrieval and reasoning, including reinforcement learning and actor--critic training that directly optimize system-level objectives through interaction.
In contrast, our work emphasizes an explicit and modular control interface based on structured sufficiency and gap prediction learned via trajectory distillation.
These directions are complementary.
The structured outputs of S2G-Judge could be incorporated into learning-based optimization as supervision, as auxiliary rewards, or as intermediate control variables when improving retrieval policies and evidence selection under practical constraints.

Second, our gap-item schema introduces a representational trade-off.
It favors stability and ease of use over expressiveness, and some questions may require richer structure than can be captured by the current fields, such as multi-entity joins, temporal constraints, or compositional relations spanning several intermediate variables.
Extending the schema, or inducing more structured intermediate programs while preserving auditability, is a promising direction for broader coverage on complex queries.

Finally, sentence-level evidence extraction introduces a compactness--recall trade-off.
While it limits the influence of redundant or distractor-heavy text and improves auditability, it may omit surrounding context that is useful for disambiguation or miss evidence that spans multiple sentences.
Moreover, our sufficiency predictions can be conservative, which in practice reduces the risk of answering from borderline evidence states, while leaving room to improve calibration so that clearly sufficient contexts are more often recognized as such.
Future work could improve calibration of sufficiency decisions, and adaptive evidence selection strategies that better balance evidence coverage and context focus.

\section*{Acknowledgments}
This work was supported by the National Natural Science
Foundation of China (No. 62376178), and the Priority Academic Program Development of Jiangsu Higher Education Institutions.

\bibliography{custom}

\appendix

\section{Additional Results}
\label{sec:additional_results}

This appendix reports supplementary results that are omitted from the main paper for space.
Unless otherwise noted, we keep the retrieval corpus and evaluation protocol identical to the main experiments, and change only the component stated in each subsection.

\subsection{Strong Reasoner Setting}
\label{sec:strong_reasoner_setting}

\begin{table}[H]
\centering
\small
\setlength{\tabcolsep}{6pt}
\begin{tabular}{l cc cc cc}
\toprule
Retriever & \multicolumn{2}{c}{TriviaQA} & \multicolumn{2}{c}{HotpotQA} & \multicolumn{2}{c}{2Wiki} \\
& EM & F1 & EM & F1 & EM & F1 \\
\midrule
BM25 & 70.7 & 79.6 & 51.0 & 65.4 & 53.6 & 61.7 \\
E5   & 72.0 & 81.1 & 51.5 & 65.3 & 53.0 & 61.3  \\
\bottomrule
\end{tabular}
\caption{S2G-RAG under a stronger reasoner setting.
We replace the answer reasoner and the sentence-level evidence extractor with GPT-4o-mini, while keeping the same retrieval.
Results are reported under BM25 and E5 on TriviaQA, HotpotQA, and 2WikiMultiHopQA.}
\label{tab:appendix_strong_reasoner}
\end{table}

Table~\ref{tab:appendix_strong_reasoner} isolates the effect of using a stronger closed-source backbone for reasoning and sentence selection.
The gains indicate that the proposed control interface remains compatible with stronger downstream components and does not require changes to the retriever or retraining the judge.
We report these results in the appendix because the main comparison in Table~\ref{tab:main_results} focuses on a matched open-source reasoner setting for fair baseline alignment.

\subsection{Query Construction Variants}
\label{sec:app_query_variants}

\begin{table}[H]
\centering
\footnotesize
\setlength{\tabcolsep}{6pt}
\resizebox{\columnwidth}{!}{%
\begin{tabular}{l c c c}
\toprule
\multicolumn{1}{l}{Variant} & EM & F1 & Correct Retrieval \\
\midrule
\multicolumn{4}{c}{\textit{Panel A: Query construction format}} \\
Free-text gap query  & 41.5 & 54.5 & 50.5 \\
Structured gap query (ours; $K{=}1$) & \textbf{43.3} & \textbf{56.5} & \textbf{52.7} \\
\midrule
$\Delta$ (Structured -- Free-text) & +1.8 & +2.0 & +2.2 \\
\midrule
\multicolumn{4}{c}{\textit{Panel B: Number of gap items used in $\mathrm{BuildQuery}$ (structured queries)}} \\
Structured query ($K{=}1$) & \textbf{43.3} & \textbf{56.5} & 52.7 \\
Structured query ($K{=}2$) & 42.3 & 55.6 & 52.4 \\
Structured query ($K{=}3$) & 43.0 & 55.9 & 52.5 \\
\bottomrule
\end{tabular}%
}
\caption{Query construction variants on HotpotQA (dev) under BM25.
Panel A compares free-text versus structured gap-based query construction.
Panel B varies the number of structured gap items $K$ used by $\mathrm{BuildQuery}$ within the structured setting.
Correct Retrieval is reported as a percentage (\%) on the final row per question ID.
All other components are kept identical.}
\label{tab:app_query_variants}
\end{table}

Table~\ref{tab:app_query_variants} analyzes the impact of query construction in the iterative retrieval loop.
In Panel A, we compare two ways of forming the next-turn query when the judge indicates missing information.
The \emph{free-text} variant appends a natural-language gap description to the original question, whereas the \emph{structured} variant constructs a query phrase from the missing-fact fields (e.g., \texttt{target} and \texttt{slot}), yielding a more consistent lexical form for BM25.
The structured variant improves EM and F1 by 1.8 and 2.0 points, and increases final-turn Correct Retrieval by 2.2 points, suggesting that structured fields provide stronger retrieval cues over multiple turns.

Panel B varies the number of structured gap items used by $\mathrm{BuildQuery}$.
Across $K\in\{1,2,3\}$, performance is stable: EM/F1 varies within 1 point and Correct Retrieval remains in a narrow band (52.4--52.7\%).
Using a single gap item ($K{=}1$) yields the best EM/F1, but the differences are modest, indicating that the method is not sensitive to the exact value of $K$ within this range.
We therefore default to $K{=}1$ as a simple configuration that keeps queries concise while achieving the strongest average accuracy.

\subsection{Effect of Judge Backbone}
\label{sec:judge_backbone}

\begin{table}[t]
\centering
\small
\setlength{\tabcolsep}{4pt}
\begin{adjustbox}{width=\columnwidth,center}
\begin{tabular}{l cc cc cc}
\toprule
\multirow{2}{*}{Judge backbone} 
& \multicolumn{2}{c}{TriviaQA} 
& \multicolumn{2}{c}{HotpotQA} 
& \multicolumn{2}{c}{2Wiki} \\
\cmidrule(lr){2-3} \cmidrule(lr){4-5} \cmidrule(lr){6-7}
& EM & F1 & EM & F1 & EM & F1 \\
\midrule
Llama-3.2-3B-Instruct & 72.0 & 77.9 & 43.3 & 56.5 & 41.7 & 48.6 \\
Qwen-3-4B-Instruct    & 70.4 & 77.4 & 44.6 & 56.0 & 42.8 & 50.2 \\
\bottomrule
\end{tabular}
\end{adjustbox}
\caption{Effect of the S2G-Judge backbone on end-to-end QA.
We swap two LoRA-finetuned judge backbones while keeping the rest of the pipeline fixed (BM25 retriever, query construction, evidence extractor, and answer reasoner: Llama-3-8B-Instruct). Both judges are trained with the same trajectory-distillation procedure.}
\label{tab:judge_backbone}
\end{table}

We study the sensitivity of S2G-RAG to the controller backbone by swapping two LoRA-finetuned S2G-Judge variants, Llama-3.2-3B-Instruct and Qwen-3-4B-Instruct, while holding all other components fixed (BM25 retriever, query construction, sentence-level evidence extraction, and Llama-3-8B-Instruct as the answer reasoner).

As shown in Table~\ref{tab:judge_backbone}, both backbones yield similar end-to-end accuracy across datasets.
Qwen-3-4B provides small gains on TriviaQA and HotpotQA EM with comparable F1, while results on 2WikiMultiHopQA are close.
Overall, the structured sufficiency-and-gap interface is compatible with different lightweight controller backbones under the same retrieval setting.

\subsection{Robustness to Retrieval Breadth}
\label{sec:robust_retrieval_breadth}

\begin{figure}[t]
  \centering
  \includegraphics[width=\columnwidth]{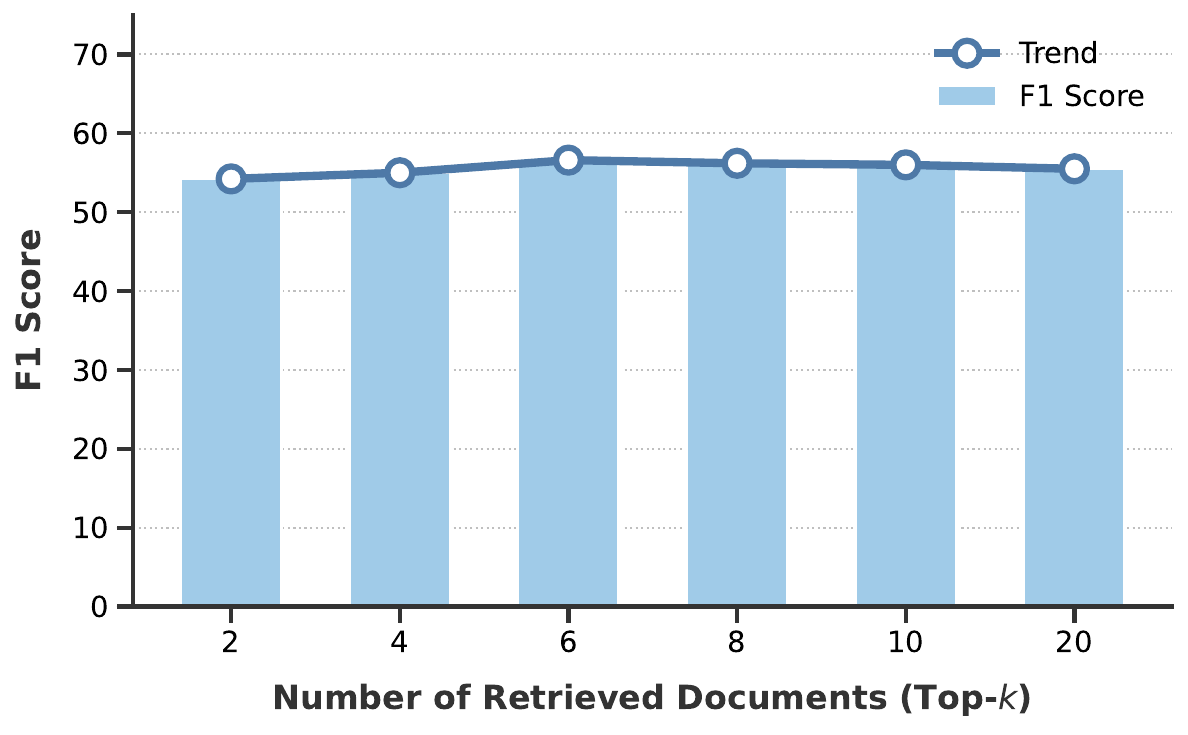}
  \caption{Robustness to per-turn retrieval breadth. We vary BM25 top-$k$ at each retrieval turn ($k \in \{2,4,6,8,10,20\}$) while keeping the judge, gap-guided query construction, evidence extractor, maximum turns $T$, and decoding settings fixed.}
  \label{fig:ctx_robust}
\end{figure}

We assess robustness to retrieval breadth by varying the number of BM25 documents retrieved per turn (top-$k$), while holding the rest of the S2G-RAG pipeline constant.
Figure~\ref{fig:ctx_robust} shows that EM/F1 remains stable across a wide range of $k$, with only modest fluctuations as retrieval breadth increases.

This behavior is largely attributable to sentence-level evidence management.
Rather than appending full documents, S2G-RAG uses a sentence-level Evidence Extractor to form a compact evidence block from the retrieved set before the next round of judging and final answering.
As $k$ grows, newly added documents are increasingly redundant or distractor-heavy, so the marginal benefit of wider retrieval tends to diminish.
By selecting a bounded set of salient sentences, the extractor limits the amount of irrelevant text that enters the accumulated evidence context, which helps keep both gap identification and answer generation stable under larger per-turn retrieval breadth.

\subsection{Effect of Evidence Extraction Cap $K_e$}
\label{sec:app_extractor_ke}

The sentence-level Evidence Extractor updates the evidence context by selecting a small set of salient sentences from the newly retrieved documents at each turn.
We study how sensitive S2G-RAG is to the extractor cap by varying $K_e \in \{2,4,6,8,10\}$ on HotpotQA (dev) under BM25, while keeping the judge, query construction, retriever settings, maximum turns, and the answer reasoner fixed.

\begin{table}[t]
\centering
\small
\setlength{\tabcolsep}{6pt}
\begin{tabular}{c cc}
\toprule
Extractor cap $K_e$ & EM & F1 \\
\midrule
2  & 43.2 & 56.5 \\
4  & 43.4 & 55.8 \\
6  & 43.3 & 56.5 \\
8  & 43.4 & 55.9 \\
10 & 43.4 & 56.0 \\
\bottomrule
\end{tabular}
\caption{Effect of the extractor cap $K_e$ (maximum number of evidence sentences returned per turn) on HotpotQA (dev) under BM25. The extractor may return fewer than $K_e$ sentences. All other components are kept identical.}
\label{tab:extractor_ke}
\end{table}

Table~\ref{tab:extractor_ke} shows that performance is stable across a wide range of caps.
EM varies within 43.2--43.4 and F1 within 55.8--56.5, with no clear monotonic trend as $K_e$ increases.
This suggests that the system is not tightly tuned to a particular extraction budget.
Intuitively, $K_e$ acts as a ceiling rather than a target: the extractor can return fewer sentences when retrieval results offer limited relevant content.
Moreover, HotpotQA instances are typically supported by a small number of key evidence sentences, so modest caps are already sufficient to preserve the essential two-hop evidence chain while limiting the inclusion of distractor text.

\subsection{Breakdown by HotpotQA Question Type}

We further analyze performance on HotpotQA by question type. Table~\ref{tab:hotpot_type_breakdown} reports results on the two standard categories, Bridge and Comparison. Our method performs well on both types, indicating that the framework is not restricted to bridge-style reasoning and also remains effective for comparison-style multi-hop QA.

\begin{table}[t]
\centering
\small
\begin{tabular}{lcc}
\toprule
Type & EM & F1 \\
\midrule
Bridge & 39.4 & 54.2 \\
Comparison & 58.6 & 67.6 \\
\bottomrule
\end{tabular}
\caption{Performance breakdown on HotpotQA by question type.}
\label{tab:hotpot_type_breakdown}
\end{table}

\section{Training Details}
\label{sec:training_details}

This appendix details the construction of turn-level supervision for S2G-Judge and the LoRA-based supervised fine-tuning procedure. Unless stated otherwise, all unspecified settings use library defaults.

\subsection{Turn-Level Supervision Construction}
\paragraph{Execution-trace collection.}
We construct supervision from multi-turn pipeline execution traces produced by the same iterative retrieval loop used at inference time (Appendix~\ref{sec:pseudocode}).
For each training question, we roll out the pipeline to a fixed maximum budget $T$ and record per-turn snapshots $x_t=(q, C_t)$, where $q$ is the original question and $C_t$ is the accumulated evidence context at turn $t$.

To assign an initial sufficiency tag for filtering, we use the gold supporting documents.
A snapshot is tagged as sufficient if the retrieved titles up to turn $t$ cover all gold supporting document titles for the instance; otherwise it is tagged as insufficient.

To keep the recorded evidence contexts representative of the pipeline’s intermediate states, trace collection is driven by an unfine-tuned judge backbone that uses the same prompt and output schema as the final S2G-Judge.
This yields evidence contexts that naturally include redundancy and distractors from multi-round retrieval and accumulation.

\paragraph{Teacher labeling.}
A stronger teacher model, prompted with our instruction template (Appendix~\ref{sec:app_teacher_prompt}), labels each snapshot under a strict context-only evidence constraint.
The teacher outputs a binary sufficiency decision together with structured gap items, based only on the information contained in $C_t$ rather than parametric knowledge.

\subsection{Supervision Filtering}
We apply lightweight filtering to improve the reliability of the distilled turn-level supervision, without introducing any gold signals into model inputs or evaluation:
\begin{itemize}
    \item \textbf{Format checking.} We validate that the teacher output follows the required schema and discard examples with malformed or non-conformant outputs.

    \item \textbf{Conflict-based filtering.} For each snapshot, we compute a coarse, retrieval-derived sufficiency tag based on whether the \emph{retrieved titles} cover the gold supporting page titles. We use this tag only as a weak sanity check to remove \emph{obvious conflicts} between the snapshot state and the teacher label (e.g., cases where the teacher appears to violate the context-only constraint, or the snapshot/tag alignment is unreliable). Importantly, this gold-derived tag is \emph{never} provided to the judge or used at inference time; it is used only to discard a small set of low-confidence supervision pairs.
\end{itemize}

\subsection{Supervision Dataset}
The final supervision file consists of per-turn examples containing the original question $q$, the accumulated evidence context $C_t$, a binary sufficiency label, and a set of structured gap items.

\paragraph{Dataset statistics.}
Table~\ref{tab:stats} summarizes the cleaned supervision set.
It contains $2{,}804$ snapshots with a relatively even turn distribution from $t{=}1$ to $t{=}4$ and approximately balanced sufficiency labels.

\begin{table}[t]
\centering
\small
\renewcommand{\arraystretch}{1.1}
\begin{tabularx}{\columnwidth}{X r}
\toprule
\multicolumn{2}{c}{\textbf{Cleaned supervision statistics}} \\
\midrule
\multicolumn{2}{l}{\textit{Turn distribution}} \\
\cmidrule(lr){1-2}
$t{=}1$ & $764$ \\
$t{=}2$ & $739$ \\
$t{=}3$ & $653$ \\
$t{=}4$ & $648$ \\
\midrule
\multicolumn{2}{l}{\textit{Sufficiency labels}} \\
\cmidrule(lr){1-2}
Sufficient   & $1{,}396$ \\
Insufficient & $1{,}408$ \\
\midrule
Total snapshots & $2{,}804$ \\
\bottomrule
\end{tabularx}
\caption{Statistics of the cleaned turn-level supervision set used to train S2G-Judge.}
\label{tab:stats}
\end{table}

\paragraph{Train--validation split.}
We randomly split the cleaned supervision set into $90\%$ training and $10\%$ validation using a fixed random seed $42$, yielding $2{,}523$ training snapshots and $281$ validation snapshots.

\subsection{LoRA Configuration}
We fine-tune S2G-Judge with LoRA~\citep{hu2022lora} for parameter-efficient adaptation.
Unless otherwise noted, we insert LoRA adapters into the attention and feed-forward projection layers in each decoder block:
\begin{itemize}
    \item \textbf{Rank} $r=16$
    \item \textbf{Scaling} $\alpha=32$
    \item \textbf{Dropout} $0.05$
    \item \textbf{Target modules} \texttt{q\_proj}, \texttt{k\_proj}, \texttt{v\_proj}, \texttt{o\_proj}, \texttt{gate\_proj}, \texttt{up\_proj}, \texttt{down\_proj}
\end{itemize}

\subsection{Supervised Fine-Tuning Setup}
We run supervised fine-tuning with TRL SFTTrainer on chat-formatted prompts.
We use the following hyperparameters:
\begin{itemize}
    \item \textbf{Epochs} $3$
    \item \textbf{Learning rate} $1\times 10^{-4}$
    \item \textbf{Gradient accumulation} $8$
    \item \textbf{Max sequence length} $2048$
    \item \textbf{Precision} FP16
    \item \textbf{Evaluation / checkpointing} evaluate every $200$ steps; save every $200$ steps; keep the latest $3$ checkpoints
\end{itemize}
After training, we save the LoRA adapter weights and the tokenizer for inference-time loading.

\section{Pseudocode for the Evaluation Pipeline}
\label{sec:pseudocode}
Algorithm~\ref{alg:eval} summarizes our evaluation pipeline for S2G-RAG.
Given a question $q$, the system iteratively maintains an accumulated evidence context $C_t$ that stores only sentence-level evidence blocks extracted from newly retrieved documents.
At each turn, the judge inspects $(q, C_t)$ and outputs a sufficiency decision $s_t$ and structured gap items $G_t$.
When $s_t$ indicates the current evidence supports answering (or when the maximum budget $T$ is reached), the reasoner produces the final prediction using only $(q, C_t)$.
Otherwise, the predicted gaps are mapped into a retrieval query, documents are retrieved, and the evidence extractor selects a compact set of sentences to append to the evidence context.

\begin{algorithm}[H]
\small
\caption{Evaluation framework of S2G-RAG.}
\label{alg:eval}
\begin{algorithmic}[1]
\Require \parbox[t]{0.90\linewidth}{
Dev set $\mathcal{S}=\{q_i\}_{i=1}^{N}$;\\
Retriever $\mathrm{Retrieve}(\cdot)$ over corpus $\mathcal{D}$;\\
Judge $\mathrm{Judge}(q, C_t)\rightarrow (s_t, G_t)$;\\
Query builder $\mathrm{BuildQuery}(q, G_t)\rightarrow \tilde q_t$;\\
Evidence extractor $\mathrm{Extract}(q, G_t, D_t)\rightarrow E_t$;\\
Reasoner $\mathrm{Reason}(q, C_t)\rightarrow \hat a$;\\
Max turns $T$.
}

\Ensure
Per-turn execution traces and final predictions.

\Procedure{Evaluate}{$\mathcal{S}$}
\For{each question $q \in \mathcal{S}$}
    \State $C_0 \leftarrow \emptyset$ \Comment{accumulated evidence memory}
    \For{$t=0$ \textbf{to} $T$}
        \State $(s_t, G_t) \leftarrow \mathrm{Judge}(q, C_t)$
        \If{$s_t=\texttt{true}$ \textbf{or} $t=T$}
            \State $\hat a \leftarrow \mathrm{Reason}(q, C_t)$
            \State Record the final trace at turn $t{+}1$
            \State \textbf{break}
        \Else
            \State $\tilde q_t \leftarrow \mathrm{BuildQuery}(q, G_t)$
            \State $D_t \leftarrow \mathrm{Retrieve}(\tilde q_t)$
            \State $E_t \leftarrow \mathrm{Extract}(q, G_t, D_t)$
            \State $C_{t+1} \leftarrow C_t \oplus E_t$
            \State Record the trace at turn $t{+}1$
        \EndIf
    \EndFor
\EndFor
\EndProcedure
\end{algorithmic}
\end{algorithm}

\section{Case Studies}
\label{sec:case_studies}
To provide qualitative insight, we report representative multi-turn traces from HotpotQA. Table~\ref{tab:case_study_rieders_swango} is a success case: the judge first predicts a relation gap about who Fredric Rieders testified against, which surfaces the correct bridge entity (Michael Swango); it then requests a missing evidence span to verify the remaining constraint (the “three consecutive life terms” and “Florence, Colorado” clue), after which it declares sufficiency and answers correctly.

Table~\ref{tab:case_study_failure_philipstown} is a failure case dominated by retrieval coverage and ambiguity. Although the trajectory identifies plausible villages in Philipstown, it never retrieves the gold supporting title (Valhalla Highlands Historic District) and instead accumulates evidence for a competing but locally supported district (Cold Spring Historic District). The judge remains insufficient throughout, but the final prediction is still incorrect due to exhausting the retrieval budget without reaching the correct supporting document.

\begin{table*}[t]
\centering
\small
\setlength{\tabcolsep}{6pt}
\renewcommand{\arraystretch}{1.15}

\begin{tabularx}{\textwidth}{@{}c Y@{}}
\toprule

\multicolumn{2}{@{}p{\textwidth}@{}}{\textbf{Question:} Who did Fredric Rieders tesify agains who had killed as many as 60 patients and recieved three consecutive life terms to be served in Florence Colorado?} \\
\multicolumn{2}{@{}p{\textwidth}@{}}{\textbf{Gold answer:} \textit{Michael Swango}. \hspace{1.25em}
\textbf{Prediction:} \textit{Michael Swango} (correct).} \\

\midrule

Turn & \makecell[l]{Trace (Gap $\rightarrow$ Query intent $\rightarrow$ Evidence $\rightarrow$ Decision/Output)} \\
\midrule

1 &
\begin{minipage}[t]{\linewidth}
\setlength{\parskip}{2pt}\setlength{\parindent}{0pt}
\textbf{Predicted gap item $G_t$.} relation; target: \textit{Fredric Rieders}; slot: \texttt{testified\_against}; desc: The person Fredric Rieders testified against. \par
\textbf{Query intent.} Identify who Rieders testified against; match the clue ``killed as many as 60 patients''. \par
\textbf{Evidence.} \textbf{[Fredric Rieders]} ``Some of his other well-known cases include \ldots\ another case is that of Michael Swango, or ``Dr.\ Death''---a serial killer who killed as many as 60 of his patients \ldots'' \par
\textbf{Decision/Output.} \texttt{insufficient}; bridge candidate: \textit{Michael Swango}.
\end{minipage}
\\

\addlinespace[4pt]

2 &
\begin{minipage}[t]{\linewidth}
\setlength{\parskip}{2pt}\setlength{\parindent}{0pt}
\textbf{Predicted gap item $G_t$.} evidence\_span; target: \textit{Michael Swango}; slot: \texttt{sentence\_length}; desc: Michael Swango's sentence length. \par
\textbf{Query intent.} Verify ``three consecutive life terms'' and the Florence, Colorado clue. \par
\textbf{Evidence.} \textbf{[Michael Swango]} ``He was sentenced in 2000 to three consecutive life terms without the possibility of parole, and is serving that sentence at the ADX Florence supermax prison near Florence, Colorado.'' \par
\textbf{Decision/Output.} \texttt{sufficient}; answer: \textit{Michael Swango}.
\end{minipage}
\\

\bottomrule
\end{tabularx}

\caption{A successful two-hop case. The judge first surfaces a bridge entity via a relation gap (Turn~1), then verifies the remaining constraint via an evidence-span gap (Turn~2).}
\label{tab:case_study_rieders_swango}
\end{table*}

\begin{table*}[t]
\centering
\small
\setlength{\tabcolsep}{6pt}
\renewcommand{\arraystretch}{1.15}

\begin{tabularx}{\textwidth}{@{}c Y@{}}
\toprule
\multicolumn{2}{@{}l@{}}{\textbf{Question:} What national historic district is located near a village in the town of Philipstown, New York?} \\
\multicolumn{2}{@{}l@{}}{\textbf{Gold answer:} \textit{Valhalla Highlands Historic District}. \hspace{1.25em}
\textbf{Prediction:} \textit{Cold Spring Historic District} (incorrect).} \\
\midrule
Turn & Trace (Gap $\rightarrow$ Query intent $\rightarrow$ Evidence $\rightarrow$ Decision/Output) \\
\midrule

1 &
\begin{minipage}[t]{\linewidth}
\setlength{\parskip}{0pt}\setlength{\parindent}{0pt}
\textbf{Predicted gap items $G_t$.}
(i) relation; target: \textit{Philipstown, New York}; slot: \texttt{\detokenize{village_name}}; desc: The name of the village in Philipstown, New York.
(ii) relation; target: \textit{village in Philipstown, New York}; slot: \texttt{\detokenize{national_historic_district}}; desc: The national historic district located near the village in the town of Philipstown, New York. \newline
\textbf{Query intent.} Identify candidate villages in Philipstown; then look for a nearby national historic district. \newline
\textbf{Evidence.} \textbf{[Cold Spring, New York]} ``Cold Spring is a village in the town of Philipstown \ldots'';
\textbf{[Nelsonville, New York]} ``Nelsonville is a \ldots village located in the town of Philipstown \ldots'';
plus many off-topic district pages. \newline
\textbf{Decision/Output.} \texttt{insufficient}.
\end{minipage}
\\

\addlinespace[4pt]

2 &
\begin{minipage}[t]{\linewidth}
\setlength{\parskip}{0pt}\setlength{\parindent}{0pt}
\textbf{Predicted gap item $G_t$.}
relation; target: \textit{Nelsonville, New York}; slot: \texttt{\detokenize{national_historic_district}}; desc: The national historic district located near Nelsonville, New York. \newline
\textbf{Query intent.} Search for a national historic district near Nelsonville (as the inferred village). \newline
\textbf{Evidence.} Only village-location support repeats; no evidence links Nelsonville to the asked district; the gold title does not appear. \newline
\textbf{Decision/Output.} \texttt{insufficient}.
\end{minipage}
\\

\addlinespace[4pt]

3 &
\begin{minipage}[t]{\linewidth}
\setlength{\parskip}{0pt}\setlength{\parindent}{0pt}
\textbf{Predicted gap items $G_t$.}
(i) relation; target: \textit{Fish and Fur Club}; slot: \texttt{\detokenize{location}}; desc: The location of the Fish and Fur Club in relation to the village in the town of Philipstown, New York.
(ii) evidence\_span; target: \textit{Nelsonville, New York}; slot: \texttt{\detokenize{national_historic_district}}; desc: A sentence explicitly stating that Nelsonville, New York is a national historic district. \newline
\textbf{Query intent.} Use nearby landmarks to triangulate the missing district signal. \newline
\textbf{Evidence.} \textbf{[Fish and Fur Club]} ``The original Fish and Fur Club building is at \ldots in Nelsonville \ldots'';
\textbf{[J. Y. Dykman Store]} ``\ldots located \ldots in the village of Nelsonville \ldots across from the Fish and Fur Club \ldots'' (useful but off-target). \newline
\textbf{Decision/Output.} \texttt{insufficient}.
\end{minipage}
\\

\addlinespace[4pt]

4 &
\begin{minipage}[t]{\linewidth}
\setlength{\parskip}{0pt}\setlength{\parindent}{0pt}
\textbf{Predicted gap items $G_t$.}
(i) relation; target: \textit{Fish and Fur Club}; slot: \texttt{\detokenize{location}}; desc: The location of the Fish and Fur Club in relation to the village in the town of Philipstown, New York.
(ii) relation; target: \textit{Fish and Fur Club}; slot: \texttt{\detokenize{national_historic_district}}; desc: The national historic district associated with the Fish and Fur Club. \newline
\textbf{Query intent.} Resolve the missing district by following auxiliary entities; fall back to the strongest surfaced village branch. \newline
\textbf{Final-turn evidence.}
\textbf{[Cold Spring, New York]}
``Cold Spring is a village in the town of Philipstown in Putnam County, New York, United States.'' \\
``The central area of the village is on the National Register of Historic Places as the Cold Spring Historic District \ldots'' \newline
\textbf{Decision/Output.} \texttt{insufficient}; predicted answer \textit{Cold Spring Historic District} (incorrect).
\end{minipage}
\\

\bottomrule
\end{tabularx}

\caption{Failure case. The trajectory identifies villages in Philipstown but never retrieves the gold supporting title \textit{Valhalla Highlands Historic District}. It instead converges on a locally supported but incorrect district (\textit{Cold Spring Historic District}) surfaced in the final turn.}
\label{tab:case_study_failure_philipstown}
\end{table*}

\section{Prompts (Judge / Teacher / Extractor)}
\label{sec:prompts}

\subsection{S2G-Judge Prompts}
\label{sec:app_judge_prompt}

\begin{table*}[t]
\centering
\small
\setlength{\tabcolsep}{6pt}
\renewcommand{\arraystretch}{1.15}
\begin{tabularx}{\textwidth}{@{}X@{}}
\toprule
\textbf{System prompt for S2G-Judge}\\
\midrule
You are a QA/RAG sufficiency judge.

Given a QUESTION and a CONTEXT (documents retrieved so far), decide whether the CONTEXT alone
contains enough information to reliably answer the QUESTION. If not, list the gap items that describe
what information is still missing.

Output format (strict):\par
- Output exactly one JSON object and nothing else.\par
- The output must have exactly two keys: "sufficient" and "gap items".\par

Schema:\par
{\par
  "sufficient": true/false,\par
  "gap items": [\par
    {
      "category": "bridge entity | attribute | relation | evidence span | other",\par
      "target": "string",\par
      "slot": "string",\par
      "description": "string"\par
    }\par
  ]\par
}

Constraint:\par
- If "sufficient" is true, then "gap items" must be an empty list [].
\\
\bottomrule
\end{tabularx}
\caption{System prompt for S2G-Judge. The judge outputs a binary sufficiency decision and, when insufficient, structured gap items.}
\label{tab:app_selector_prompts}
\end{table*}

\subsection{Teacher Prompt}
\label{sec:app_teacher_prompt}

\begin{table*}[t]
\centering
\small
\setlength{\tabcolsep}{6pt}
\renewcommand{\arraystretch}{1.15}
\begin{tabularx}{\textwidth}{@{}X@{}}
\toprule
\textbf{Teacher prompt (Part I: task definition and labeling rules)}\\
\midrule
You are a QA/RAG sufficiency judge.

You will be given:\par
- a QUESTION\par
- the current-round CONTEXT (what the QA system has retrieved so far)\par

Your tasks:

1. Decide whether the given CONTEXT alone contains enough information to reliably answer the QUESTION.\par
- Answer using the boolean field "sufficient".\par

IMPORTANT:\par
- You MUST base your decision ONLY on the CONTEXT.\par
- Even if you personally know the correct answer from world knowledge or training data, if the CONTEXT does not clearly support a correct answer, you MUST set "sufficient": 
false.\par
- Only when the CONTEXT itself already provides enough explicit evidence to justify a correct answer, you may set "sufficient": true.

2. If the information is NOT sufficient (i.e., "sufficient": false), list the gap items needed to answer the QUESTION.\par

Use the field \texttt{\detokenize{gap_items}}: a list of 1--3 objects, each with:\par
- "category": one of ["bridge entity","attribute","relation","evidence span","other"]\par
- "target": a short string naming the entity or concept this gap is about\par
- "slot": a coarse type of the missing fact (e.g., alias, name, nationality, birth place, location)\par
- "description": a short English phrase describing the missing information

Guidelines:\par
- Only mark gap items as missing if they are NOT explicitly stated in the CONTEXT.\par
- Think about the reasoning chain and identify which links are still missing.\par
- Avoid vague statements like "need more information" without specifying what is missing.

Surface forms and aliases:\par
- If QUESTION uses one name (A) and CONTEXT introduces an alias (B), create ONE gap item with target A / B and descriptions that mention both forms.\par
- Do NOT create two separate gap items for the same underlying relation just due to surface forms.

3. If the information IS sufficient (i.e., "sufficient": true), set \texttt{\detokenize{gap_items}}: [].
\\
\bottomrule
\end{tabularx}
\caption{Teacher prompt used to label per-turn snapshots for judge training (Part I).}
\label{tab:teacher_prompt_part1_clean}
\end{table*}

\begin{table*}[t]
\centering
\small
\setlength{\tabcolsep}{6pt}
\renewcommand{\arraystretch}{1.15}
\begin{tabularx}{\textwidth}{@{}X@{}}
\toprule
\textbf{Teacher prompt (Part II: output format and example)}\\
\midrule
You MUST respond with a single JSON object and NOTHING else.

The JSON MUST have exactly the following shape:

\hspace*{1.2em}"sufficient": true/false,\\
\hspace*{1.2em}"\detokenize{gap items}": [\\
\hspace*{2.4em}\{ "category": "...", "target": "...", "slot": "...", "description": "..." \},\\
\hspace*{2.4em}...\\
\hspace*{1.2em}]\\

Do NOT add any extra keys.\\
Do NOT add explanations outside the JSON.

EXAMPLE (for style ONLY; alias is already present in CONTEXT)

QUESTION: "What nationality was A's wife?"\\
CONTEXT: (... includes "A is better known as B", but does NOT mention wife's name or nationality)\\
Output:

\hspace*{1.2em}"sufficient": false,\\
\hspace*{1.2em}"\detokenize{gap items}": [\\
\hspace*{2.4em}\{ "category": "relation", "target": "A / B", "slot": "spouse name",\\
\hspace*{4.8em}"description": "The name of A (B)'s wife." \},\\
\hspace*{2.4em}\{ "category": "attribute", "target": "A (B)'s wife", "slot": "nationality",\\
\hspace*{4.8em}"description": "The nationality of A (B)'s wife." \}\\
\hspace*{1.2em}]\\

QUESTION

<q>

CONTEXT (this is the ONLY evidence you may use)

<ctx>

Now output ONLY the JSON object:
\\
\bottomrule
\end{tabularx}
\caption{Teacher prompt used to label per-turn snapshots for judge training (Part II).}
\label{tab:teacher_prompt_part2_clean}
\end{table*}

\subsection{Sentence-level Evidence Extractor Prompts}
\label{sec:app_selector_prompt}

\begin{table*}[t]
\centering
\small
\setlength{\tabcolsep}{6pt}
\renewcommand{\arraystretch}{1.15}
\begin{tabularx}{\textwidth}{@{}X@{}}
\toprule
\textbf{System prompt (Evidence Selector)}\\
\midrule
You are a sentence-level evidence selector for a multi-hop RAG system.\par

You will receive:\par
1. an ORIGINAL QUESTION,\par
2. MISSING FACTS that describe what information is still missing,\par
3. a numbered list of SENTENCES from retrieved documents.\par

Your task is to select the sentence ids that maximize answerability for the ORIGINAL QUESTION.\par

Selection policy:\par
1. First prioritize sentences that fill the MISSING FACTS, especially bridge entities, attributes, relations, and evidence spans needed for the next hop.\par
2. Then prioritize sentences that directly support the final answer to the ORIGINAL QUESTION.\par
3. Prefer sentences that are self-contained and explicit:\par
   - they mention the key entity, relation, attribute, date, number, or answer-bearing fact;\par
   - they remain understandable when extracted alone.\par
4. If a selected sentence depends on nearby context to be understandable or useful, include the minimal additional sentence(s) needed to preserve that context.\par
5. Do not infer, rewrite, paraphrase, or generate evidence text. Only return ids from the provided list.\par
6. If no sentence is useful, return an empty list.\par

Output format (strict):\par
Return exactly one JSON object and nothing else:\par
{"evidence global ids": [1, 5, 7]}\par

Constraints:\par
- "evidence global ids" must be a JSON array of integers.\par
- Select at most K sentences, where K is given in the user message.\par
- Only use ids that appear in the numbered sentence list.\par
- Do not repeat ids.
\\
\bottomrule
\end{tabularx}
\caption{System prompt for the sentence-level Evidence Extractor. The extractor outputs sentence indices (pointers) rather than rewriting evidence text.}
\label{tab:app_selector_prompts}
\end{table*}

\end{document}